\documentstyle[11pt,epsfig]{article}
\def\baselinestretch{1.3}
\newcommand{\ba}{\begin{array}}
\newcommand{\ea}{\end{array}}
\newcommand{\bd}{\begin{displaymath}}
\newcommand{\ed}{\end{displaymath}}
\newcommand{\be}{\begin{equation}}
\newcommand{\ee}{\end{equation}}
\newcommand{\bea}{\begin{eqnarray}}
\newcommand{\eea}{\end{eqnarray}}
\newcommand{\ra}{\rightarrow}

\def\gtap{\raisebox{-.4ex}{\rlap{$\sim$}}
\raisebox{.4ex}{$>$}}

\def\q2 {q^2}

\def\slep {\tilde{l}}
\def\sq {\tilde{q}}
\def\msl {m_{\slep}}
\def\msq {m_{\sq}}

\def\miset {E_T \!\!\!\!\!\!\!/ ~~}

\begin{document}
\begin{flushright}
{\large MRI-P-010903}\\
{hep-ph/0109071}\\
\end{flushright}

\begin{center}
{\Large\bf Signals of neutralinos and charginos from 
gauge boson fusion at the CERN Large Hadron Collider}\\[20mm]
Anindya Datta \footnote{E-mail: anindya@mri.ernet.in},
Partha Konar \footnote{E-mail: konar@mri.ernet.in}   and
Biswarup Mukhopadhyaya \footnote{E-mail: biswarup@mri.ernet.in}\\
{\em Harish-Chandra Research Institute,\\
Chhatnag Road, Jhusi, Allahabad - 211 019, India},
\end{center}

\vskip 20pt
\begin{abstract}
  
  We point out that interesting signals of the non-strongly
  interacting sector of the supersymmetric standard model arise from
  the production of charginos and neutralinos via vector boson fusion
  (VBF) at the Large Hadron Collider (LHC). In particular, if R-parity
  is violated, the hadronically quiet signals of charginos and
  neutralinos through direct production get considerably suppressed.
  We show that in such cases, the VBF channel can be useful in
  identifying this sector through clean and background-free final
  states.

\end{abstract}

\vskip 1 true cm

\setcounter{footnote}{0}

\def\baselinestretch{1.8}

\section{Introduction}

It is well-established that vector boson fusion (VBF) at
a high-energy, high luminosity hadron collider provides a useful
channel for the identification and detailed study of the Higgs boson
\cite{higgs_vbf}. This technique is applicable to both a standard
Higgs and one occuring in the supersymmetric (SUSY) extension of the
standard model \cite{mssm_higgs_vbf} .  The feasibility of tagging the
two forward jets from which the gauge bosons originate, with no
hadronic activity in the rapidity interval between them, gives added
incentive to this search strategy at the Large Hadron Collider (LHC)
at CERN. In this paper, we want to emphasize that it may be worthwhile
to use the VBF channel to also look for signatures of 
the chargino-neutralino sector of the
supersymmetric standard model, at least in certain scenarios.

Let us briefly recall the generally accepted strategy for identifying
modes of direct production and decay for charginos and neutralinos at
hadron colliders. These modes are important for an understanding of
the superparticle spectrum, especially in the non-strongly interacting
sector of the theory. The most conspicuous signal is expected from
direct production of a $\chi_1^{\pm} \chi_2^0$ pair (i.e. the lighter
chargino together with the second lightest neutralino).  In the
minimal SUSY standard model (MSSM) \cite{susy} with conserved R-parity
(defined as $R = (-)^{(3B + L + 2S)}$), the $\chi_2^0$ and
$\chi_1^{\pm}$ have substantial branching ratios for leptonic decays
in the channels $\chi_2^0 \longrightarrow \chi_1^0 l^{+}l^{-}$ and
$\chi_1^{\pm} \longrightarrow \chi_1^0 l^{\pm} \nu_l \;(\bar{\nu}_l)$
respectively.  Thus one has the so-called `hadronically quiet'
$trilepton~+~\miset$ signal. Such a signal can easily be made free
from standard model backgrounds, and it bears an unambiguous stamp of
the non-strongly interacting sector of the MSSM.  Detailed estimates
of such signals, including QCD corrections, exist in the literature
\cite{trilep}.  In addition, it has also been argued that dilepton
signals, of both like and unlike signs, ensue when one of the three
leptons mentioned above escapes undetected, thereby giving additional
types of easily distinguishable events \cite{trilep_more}.

However, the leptonic decay channels might be relatively suppressed in
certain cases. For example, as we shall see afterwards, $\chi_2^0$ and
$\chi_1^{\pm}$ may decay directly into leptons and jets in an R-parity
violating scenario \cite{rpv_susy}. Since R-parity is not necessitated
by any inherent symmetry of the theory, it is perfectly natural to
envision such a situation.  Under such circumstances, while the
trilepton signals suffer from a suppression, the final states obtained
from R-violating decays of charginos and neutralinos may be difficult
to disentangle from signals of the strongly interacting sector,
namely, gluinos and squarks. It is for such situations that we point
out the usefulness of the VBF channel in isolating charginos and
neutralinos through clearly identifiable and background-free events.

In section 2, we discuss neutralino and chargino decays in an
R-parity violating model, and demonstrate that in certain cases
the MSSM decays can be suppressed. The signals for these
cases in the VBF channels, together with the event selection criteria,
are discussed in section 3. We wind up with some
concluding remarks in section 4.

\section{An R-parity violating scenario: chargino and neutralino
decays, and what happens to the direct signals}

In  general,  R-parity violation in a SUSY model 
implies that the MSSM superpotential, written
in terms of the quark, lepton and Higgs superfields, gets
augmented by the following terms \cite{barger}:

\begin{equation}
W_{\not R} = \lambda_{ijk} {\hat L}_i {\hat L}_j {\hat E}_k^c +
\lambda_{ijk}' {\hat L}_i {\hat Q}_j {\hat D}_k^c +
\lambda_{ijk}''{\hat U}_i^c {\hat D}_j^c {\hat D}_k^c + \epsilon_i {\hat
L}_i {\hat H}_2
\end{equation}

The non-observation of proton decay leads to the further assumption 
that only L (through the $\lambda$, $\lambda'$
or $\epsilon$-type couplings) or B (through the $\lambda''$-type
couplings) can be violated at a time. We demonstrate
our main point in a simplified scenario where only {\em one}
$\lambda'$-type coupling is present. We shall comment on the
other types of L-violating couplings in the last section.

In presence of a $\lambda'$-type term in the superpotential,
the following additional decay modes become available to
neutralinos and charginos:

\begin{eqnarray}
\chi^0 \longrightarrow  l^{\pm}q \bar{q}'\\
\chi^{\pm} \longrightarrow  l^{\pm}q \bar{q}
\end{eqnarray}

\noindent
the branching ratios depending on the strength of the R-violating
interaction. In addition, of course, final states with neutrinos
replacing leptons are also possible.

Our numerical estimates are done by assuming $\lambda_{221}'$
to be the only non-zero L-violating coupling. Throughout this study,
we have fixed slepton and squark masses at 200 and 300 GeV
respectively. For such a squark mass,
the maximum allowed value for $\lambda_{221}'$ is 0.54 \cite{allanach}.
We use values well within this limit to compute the branching
ratios for the decays mentioned in (2) and (3).
Gaugino mass unification at a high energy scale has been
assumed, so that all masses and mixing angles in the
chargino-neutralino sector are fixed when
we specify the SU(2) gaugino mass $M_2$, the Higgsino
mass parameter $\mu$, and $\tan \beta$, the ratio of the vacuum
expectation values (vev) of the two Higgs doublets. We have {\em not}
restricted ourselves by any further assumption
about high scale physics (such as a supergravity framework), and thus
the squark and slepton masses (and also $\mu$) have been
treated essentially as free parameters.

\begin{figure}
\centerline{
\epsfxsize= 7 cm\epsfysize=7.0cm
                     \epsfbox{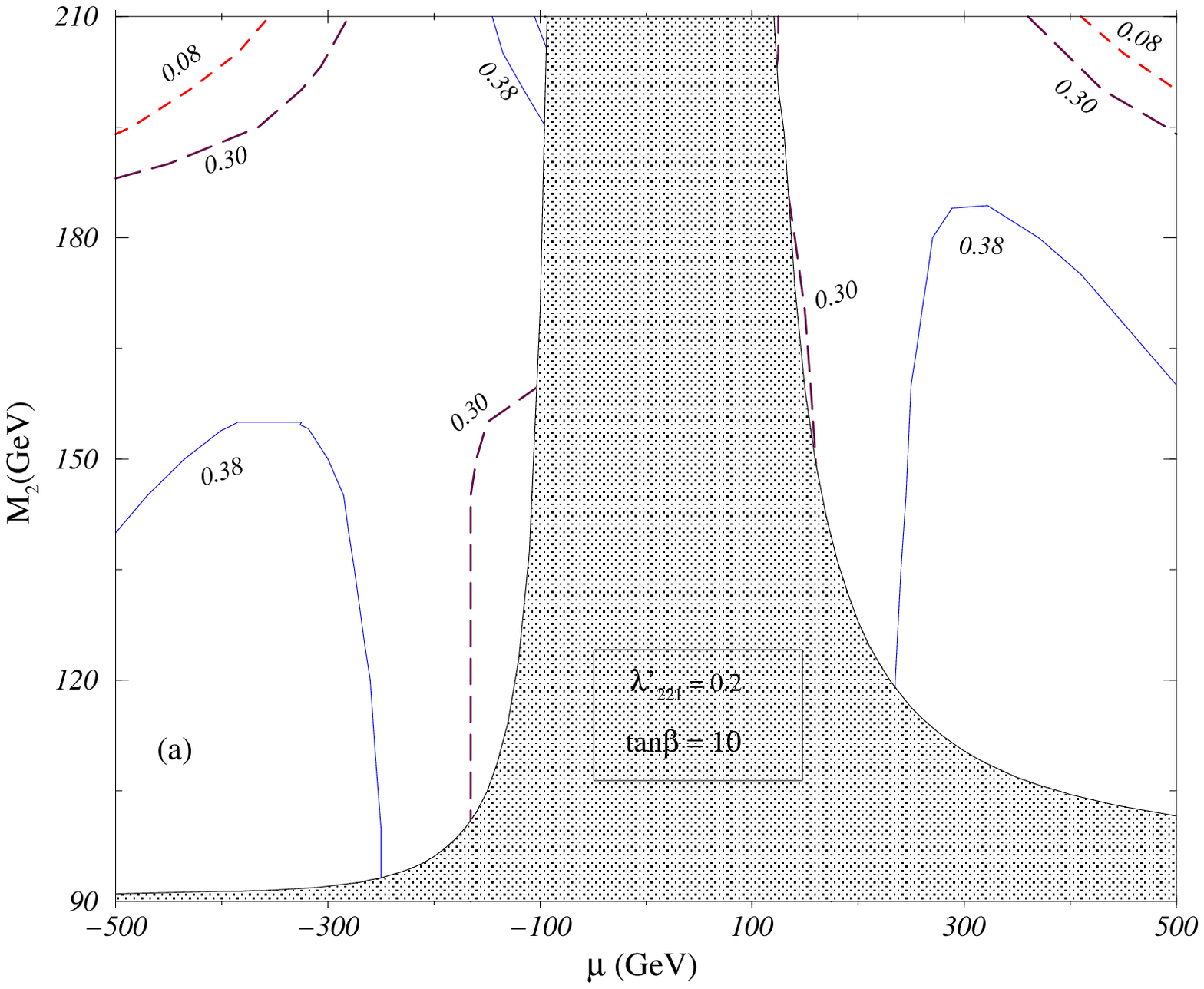}
        \hspace*{.7cm}
\epsfxsize=7.0 cm\epsfysize=7.0cm
                     \epsfbox{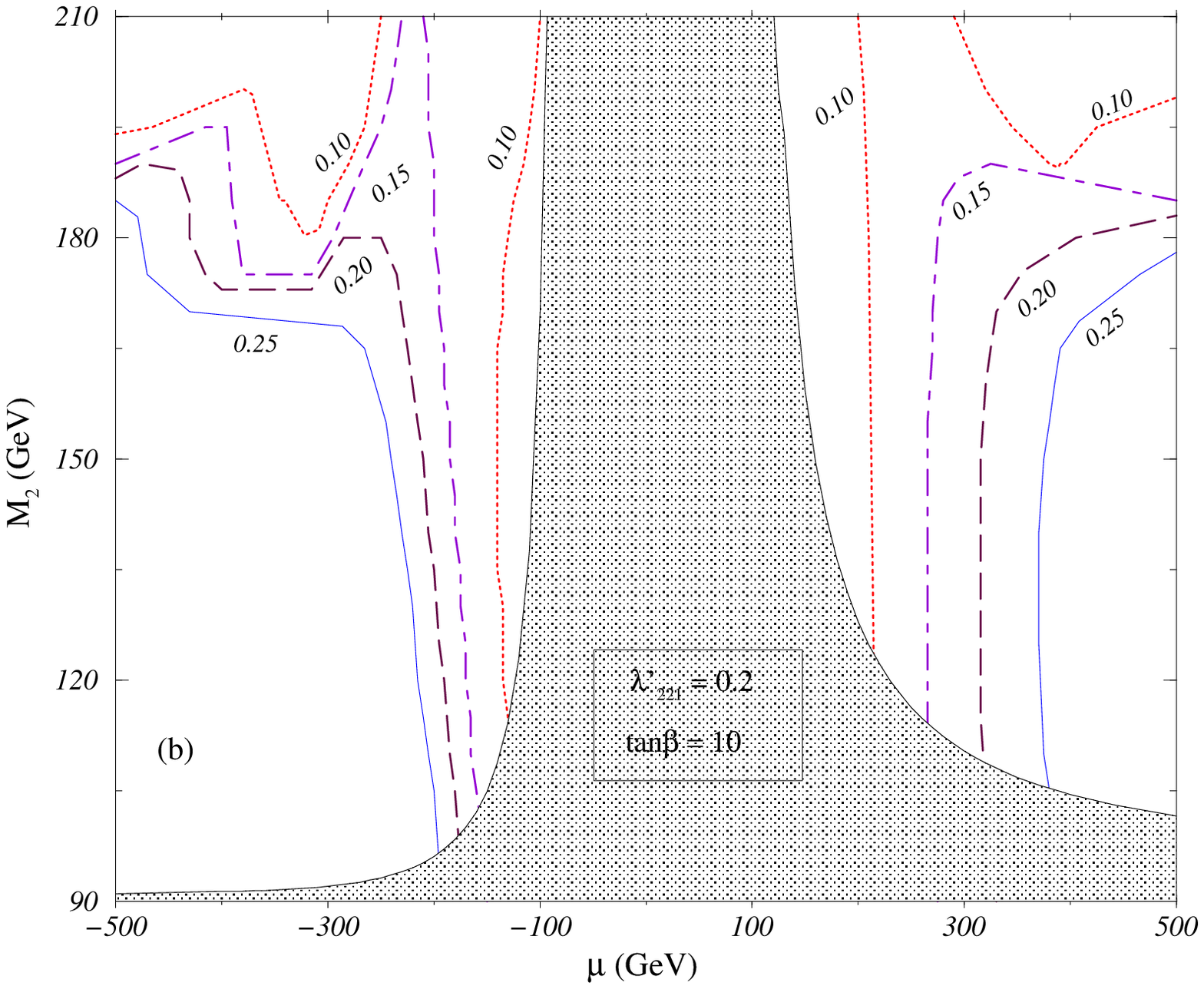}
}

\caption{{\it Contours of constant branching ratios for the decays
 (a) $\chi_2^0 \ra l^\pm q \bar q'$ and (b) $\chi^\pm \ra l^\pm q \bar q$
in $M_2 - \mu$ plane, with $\lambda '_{221} = 0.2$ and $\tan \beta = 10$.
 We have
assumed $\msq = 300 ~GeV$, $\msl = 200 ~GeV$. The shaded region is disallowed 
from LEP data.
    }}
\label{fig1}
\end{figure}

\begin{figure}
\centerline{
\epsfxsize= 7 cm\epsfysize=7.0cm
                     \epsfbox{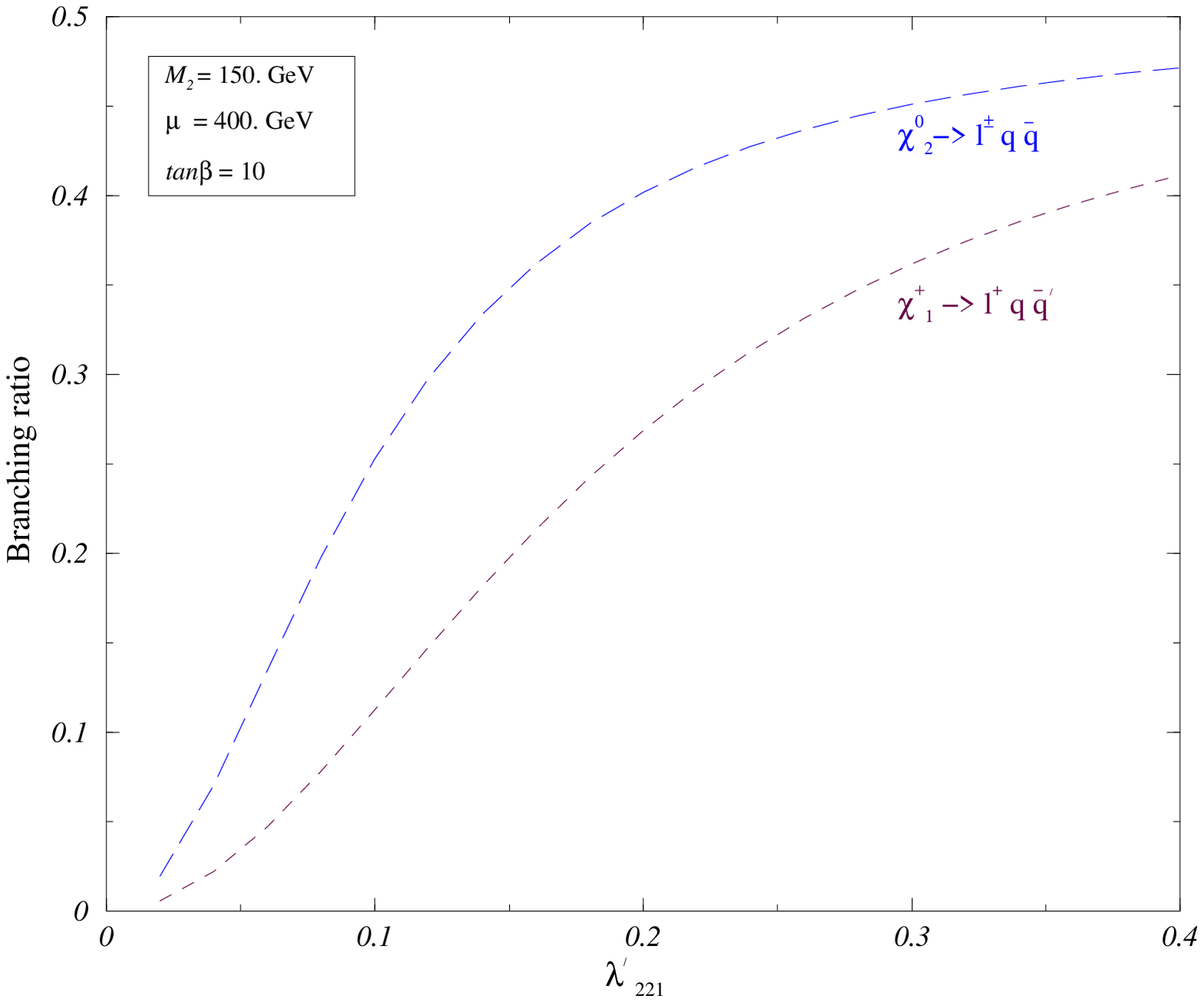}
        \hspace*{.7cm}
\epsfxsize=7.0 cm\epsfysize=7.0cm
                     \epsfbox{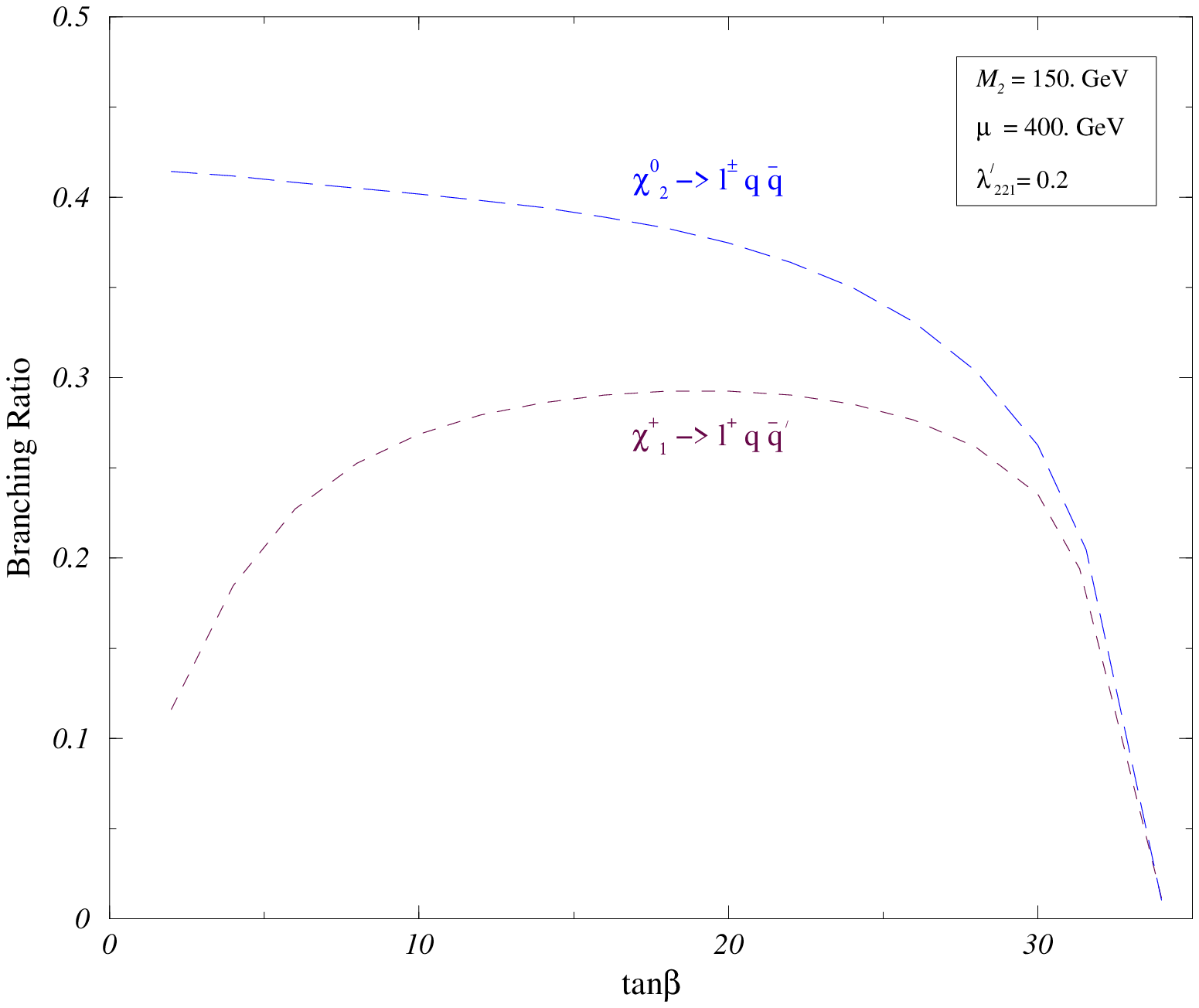}
}

\caption{{\it Variation of the branching ratios for the R-parity violating 
decays of $\chi_2^0$ and $\chi_1^{\pm}$ with (a)  the coupling 
$\lambda '_{221}$ and (b) $\tan \beta$.
    }}
\label{fig2}
\end{figure}

Some branching ratio contours for such decays for $\chi_2^0$ and
$\chi_1^+$ in the $\mu$-$M_2$ parameter space are shown in figures
\ref{fig1}(a) and \ref{fig1}(b), for $\lambda_{221}' = 0.2$ and $\tan
\beta = 10$.  The regions constrained by LEP data \cite{lep_rp} are
also shown in the same figures. We supplement these by figures
\ref{fig2}(a) and \ref{fig2}(b) , which show the variation of the same
branching ratios with $\lambda_{221}'$ and $\tan \beta$. The dip in
the curves from $\tan \beta \simeq 30$ onwards is due to the lowering
of the lighter stau mass to such a level that the tau-stau
(neutrino-stau) decay channel opens up for the $\chi_2^0$
($\chi_1^{\pm}$). Similarly, for low $\tan \beta$, the couplings
(which are functions of the chargino and neutralino mixing elements)
driving the MSSM chargino decays undergo a rise. This in turn explains the
relative insignificance of the R-parity violating widths. The
numerical results presented here use $\tan \beta = 10$, which
corresponds to the plateau in figure \ref{fig2}b.

In addition to what is shown is equation (3), the chargino will also
have an R-violating decay mode into a neutrino, driven by the same
coupling.  As for the neutralino, its Majorana character implies the
existence of four R-violating decay modes altogether. Taking all of
these into account, the contours shown in the figures imply
considerable suppression of the MSSM decay channels for $\chi_2^0$ and
$\chi_1^{\pm}$ over a substantial region of the allowed parameter
space. On the whole, for $\mu = 400~GeV$, $M_{2} = 150~GeV$ and $\tan
\beta = 10$, the combined trilepton branching ratio for a $\chi_2^0
\chi_1^{\pm}$ pair gets reduced by about one order for $\lambda_{221}'
 = 0.2$, and by a factor of nearly 50 for $\lambda_{221}' = 0.4$.
The consequent dilution of the trilepton signals at the LHC may cost
one dearly in the exclusive search for a chargino-neutralino pair. On
the other hand, the modes available in this case lead to
$lepton\; +\; jets\; + \;\miset$ and $dilepton \;+\; jets$ events.  Similar
final states are also possible from cascade decays of squarks and
gluinos which are produced much more copiously at hadron colliders.
Estimates of such ($n\;jets + m\;leptons + \miset$) event rates
already exist for both the Tevatron and the LHC \cite{susy_lhc_tev},
and cascades originating from coloured superparticles are likely to
dominate there.  Therefore, the exclusive production modes in the
chargino-neutralino sector, which are so clean and distinctive at LHC
for MSSM, lose much of their cleanliness over a sizable region of the
allowed parameter space in a scenario with broken R-parity.

In cases like the above, it is hardly necessary to emphasize the
importance of new production channels in unravelling the non-strongly
interacting sector of SUSY, which is so essential to obtain a clear
picture about the particle spectrum and the different interactions.
One direction to explore in this connection is the production of
charginos and neutralinos through gauge boson fusion at the LHC.  We
take this up in the next section.

\section{Neutralino and chargino production via vector boson fusion}

The vector boson fusion channel has so far been studied mainly
with a view to Higgs searches. The salient features of this
class of processes are \cite{rain_thesis}:

\begin{itemize}
\item Two highly energetic forward jets (originating from
the two quarks from which the gauge bosons are emitted) moving in
opposite directions, with the
pseudorapidity ($\eta$) of each jet peaking in the region $\eta~=~3-4$.

\item Decay products of the particle(s) produced via VBF lying in the
rapidity interval between the two forward jets.

\item Suppression of hadronic activity (arising from coloured particle 
exchange) in the rapidity interval between the forward jets.
\end{itemize}

If neutralinos and charginos are also produced via VBF,
then the above event characteristics enable us to eliminate
similar effects originating from gluinos
and squarks. This can be achieved by demanding a high ($\gtap~~650~GeV$)
invariant mass for the
forward jet pair. Consequently,
when R-parity violating decays for $\chi_2^0$ and
$\chi_1^{\pm}$ dominate, the centrally produced charginos and
neutralinos should give rise to events of the type

\begin{center}
\em Like- or Unlike-Sign~ Dileptons $+$ $\ge$ $2$  Jets
\end{center}

\noindent
in the rapidity interval between the two high invariant mass forward jets.
With our particular choice of the $\lambda'$-coupling, the leptons
actually turn out to be muons.
Such events are largely free from standard model backgrounds, too,
especially after applying the cuts that we shall discuss shortly.

Signals of the above type can be obtained from $\chi_i^0 \chi_j^0$,
$\chi_i^{\pm} \chi_j^{\pm}$, $\chi_i^{+} \chi_j^{-}$ as well as
$\chi_i^{0} \chi_j^{\pm}$. Of these, the largest contributions come
from $\chi_1^{+} \chi_1^{-}$ and $\chi_2^{0} \,\chi_1^{\pm}$, so long as
one adheres to a gaugino mass unification scheme.  Out of the large
number of diagrams leading to final states mentioned above, only those
involving $W$, $Z$ and $\gamma$-fusion contribute to events with the
characteristics laid down here. We have numerically calculated 
all the helicity
amplitudes (for $pp$ collision with $\sqrt{s}~=~14~TeV$ corresponding
to these diagrams using the HELAS package \cite{helas}. CTEQ4L
distributions \cite{cteq} have been used in our parton level Monte
Carlo calculation, with the factorization scale set at the sum of the
masses of the neutralino/chargino pair-produced via VBF.

It should be noted that the signals under consideration here can come
not only from direct R-parity violating decays of heavier neutralinos
and charginos but also from, say, $\chi_2^{0}$ or $\chi_1^{\pm}$
decaying hadronically through MSSM interactions to the $\chi_1^{0}$, and
the latter decaying through $\lambda'$-type couplings. Such cascade
processes are included in our calculation; they are especially
important in cases where the R-violating couplings are relatively
small. In such cases, MSSM interactions override them in $\chi_2^{0}$ or
$\chi_1^{\pm}$ decays, but their presence is reflected in the 
decay channels of the lightest neutralino.

The events have been required to pass the following cuts:

\begin{itemize}
\item A minimum invariant mass of $650~GeV$ on the two
forward jets ($j_1, j_2)$.

\item  $2 \le |\eta| \le 5$ for each forward jet, with
$\eta(j_1)\; \eta(j_2) < 0$.

\item $|\Delta \eta_{j_{1}j_{2}}| \ge 4$.

\item $E_{T} \ge 15~GeV$ for all jets.

\item The central jets and leptons to lie in the rapidity interval between
 $j_1$ and $j_2$ .

\item $\Delta R(l,j_{1(2)}) \ge 0.6$.

\item $\Delta R(central~jet, j_{1(2)}) \ge 0.6$.

\item $p_{T} \ge 10~GeV$ for the central muons.

\item The dilepton invariant mass lying outside the region
 $m_Z ~ \pm ~ 15~GeV$ for unlike-sign dileptons.

\item Total missing $E_T \le 10 ~GeV$.
\end{itemize}

The rapidity and isolation cuts separate the central events arising
from neutralino and chargino decays. The large invariant mass demanded
of the forward jets sets the signal events apart from potential
backgrounds from squarks and gluinos. This combination of cuts
eliminates practically all standard model backgrounds as well as those
arising in R-conserving MSSM. In the case of unlike-sign leptons, the
invariant mass cut on the lepton pair takes care of faking by a
$Z$-boson.  And finally, an upper limit set on the missing transverse
momentum (obtained after applying a Gaussian smear \cite{smear} on the
momenta of the final-state partons and leptons) is expected to take
care of such backgrounds as those from a $t \bar{t}$ pairs, with
semileptonic b-decays giving rise to like-sign dileptons. For unlike-sign
dileptons, the two b's from a top-antitop pair can also in principle be
a potential source of backgrounds. A stiff isolation cut between the
leptons and jets in the central region is one cure for this problem;
however, the signal strength then gets reduced by upto a factor of 10.     
Instead of applying such a cut, we have noticed that
the upper limit set on the missing $E_T$ comes to our rescue in
reducing this background. Since the leptons and neutrinos have similar
$E_T$ distributions, those background events which pass the lepton $E_T$ 
cuts set by us mostly carry large missing $E_T$ as well, and thus they
get eliminated by the $\miset$-cut. Thus the demand that our final 
state particles are `visible' makes the signal background-free.

\begin{figure}
\centerline{
\epsfxsize= 7 cm\epsfysize=7.0cm
                     \epsfbox{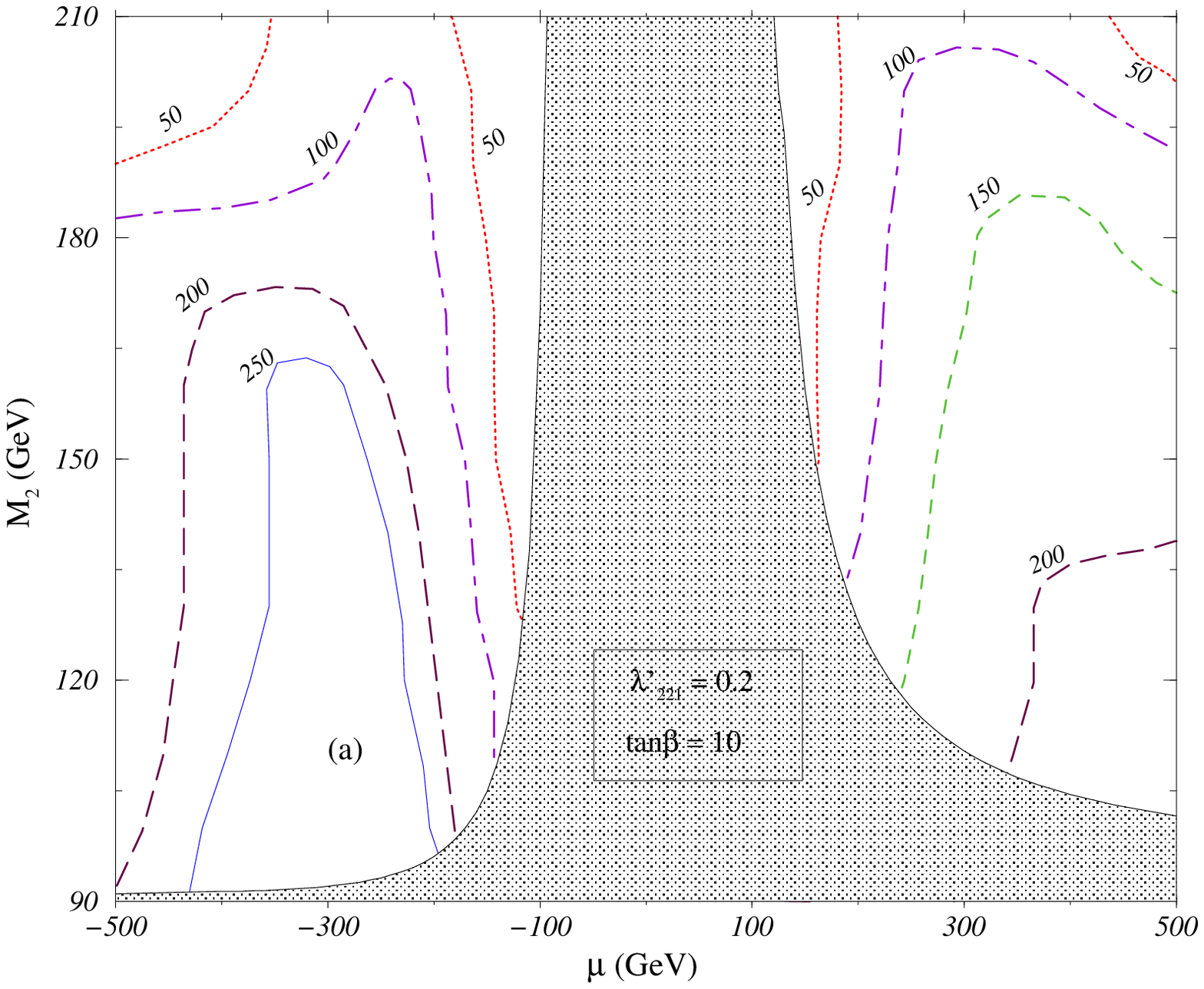}
        \hspace*{.7cm}
\epsfxsize=7.0 cm\epsfysize=7.0cm
                     \epsfbox{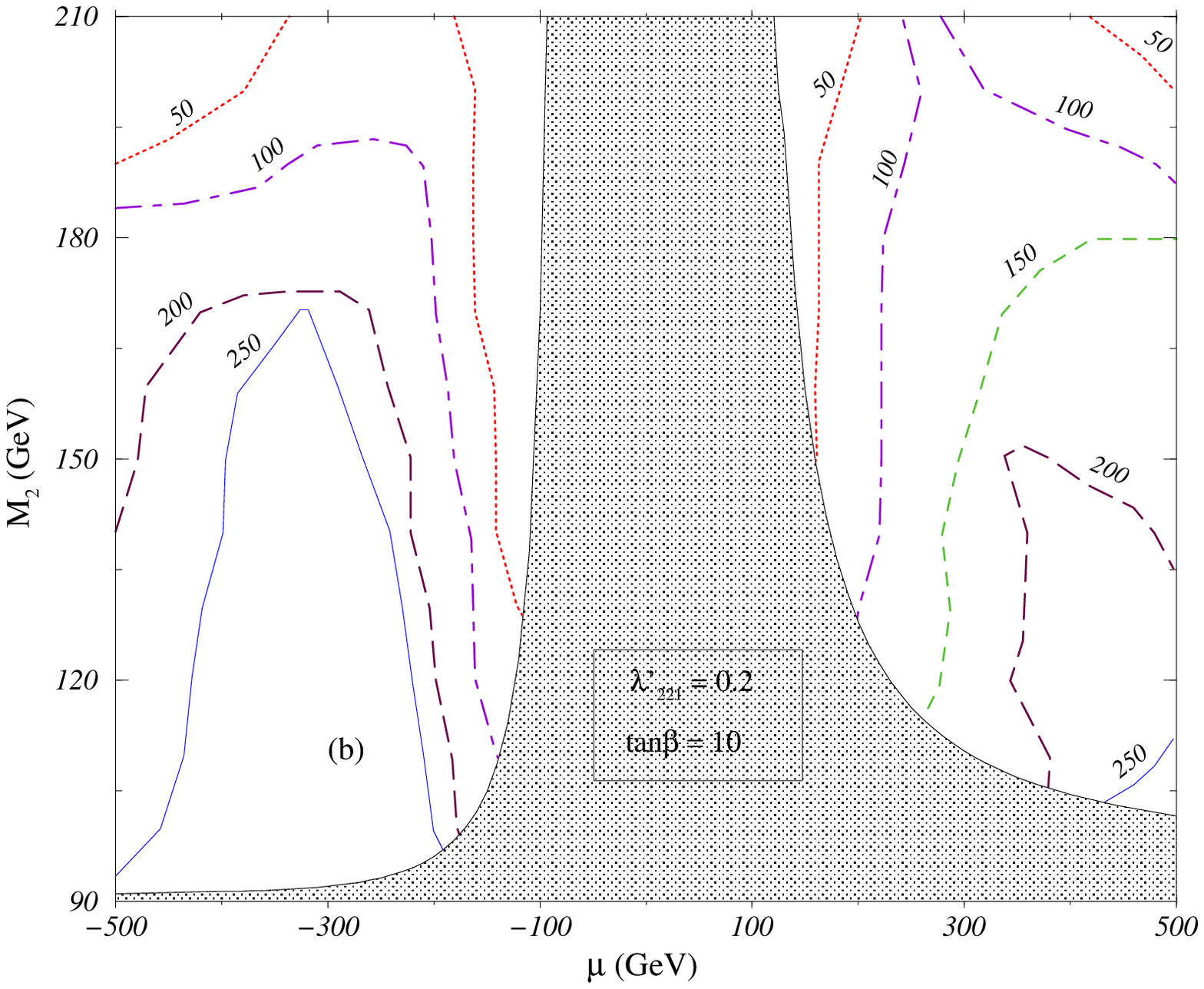}
}

\caption{{Event contours for (a) like-sign dilepton (LSD) production
 and (b) unlike-sign dilepton (USD) production in $M_2 - \mu$ plane, 
with $\lambda '_{221} = 0.2$ and $\tan \beta = 10$. Other features are same as 
in fig. \ref{fig1}.
    }}
\label{fig3}
\end{figure}

In figures \ref{fig3}(a) and \ref{fig3}(b), we present the event contours
for both like-and unlike-sign dileptons (LSD,USD), for $\lambda_{221}'
= 0.2$ and $\tan \beta = 10$. The event rates have been calculated
for an integrated luminosity of $100~fb^{-1}$. It is found that as far
as USD events are concerned, the dominant contributions come from
$\chi_1^{+}\chi_1^{-}$ production and R-parity violating decay of each
of them.  Also, the $\chi_1^{\pm}\chi_2^{0}$ production channel
contributes appreciably.  For LSD, on the other hand, a
$\chi_1^{+}\chi_1^{-}$ pair contributes only if at least one chargino
has an R-conserving decay into a $\chi_1^0$ which subsequently gives
one lepton via the $\lambda'$-type coupling.  In any case, one can
easily have upto about 250 events of both USD and LSD types surviving
the cuts for the choice of parameters shown here. This includes
regions in the parameters space for both positive and negative values
of $\mu$; it may be noted here that the region of our interest is
compatible with the SUSY solution recently measured excess in muon
anomalous magnetic moments, so long as $\mu$ is positive \cite{pran}.

\begin{figure}
\centerline{
\epsfxsize= 7 cm\epsfysize=7.0cm
                     \epsfbox{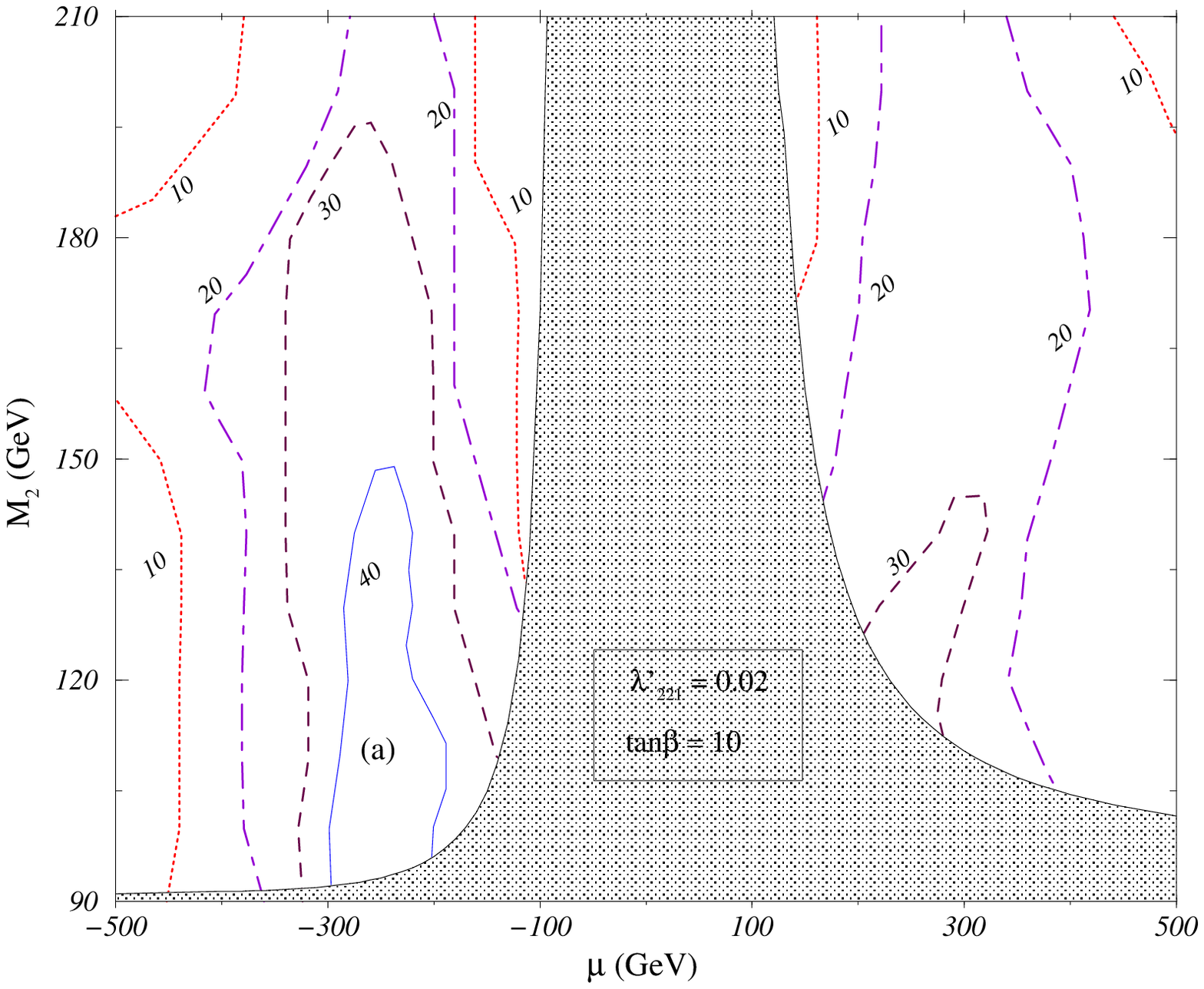}
        \hspace*{.7cm}
\epsfxsize=7.0 cm\epsfysize=7.0cm
                     \epsfbox{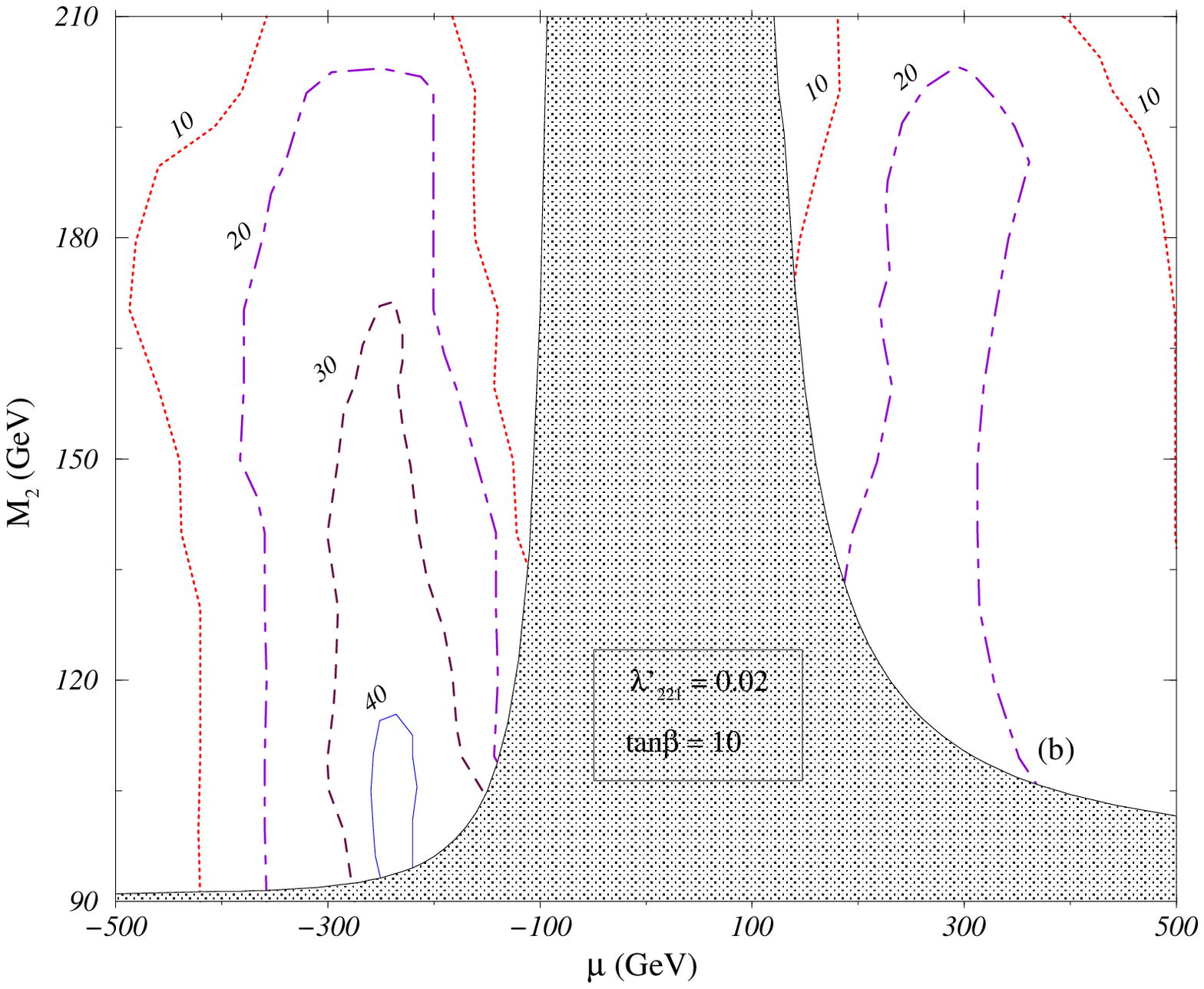}
}

\caption{{\it Event contours for (a) like-sign dilepton (LSD) 
production
and (b) unlike-sign dilepton (USD) production in $M_2 - \mu$ plane, 
with $\lambda '_{221} = 0.02$ and $\tan \beta = 10$. Other features are same as 
in fig. \ref{fig1}.
    }}
\label{fig4}
\end{figure}

Signals of the said type can be measurable even for much smaller
values of $\lambda_{221}'$. For example, if it has a value of $0.02$,
then the R-parity violating decays of $\chi_1^{\pm}$ and $\chi_2^{0}$
are severely suppressed. However, this is where the two-stage decays,
boosted by the 100\% branching ratio for R-violating $\chi_1^{0}$
decay, come to be of use. As is shown in figures \ref{fig4}(a) and
\ref{fig4}(b), one can expect up to about 40 events of both USD and LSD
types with the given integrated luminosity even with such a small
value of the R-parity violating interaction. This leads us to the
conclusion that such signals are worthwhile to investigate at the LHC
even for a situation where R-parity violation is miniscule.

We have presented our results in the $\mu - M_2$ plane. From figure 4 it is
evident that for $\lambda'_{221} =$ 0.02, one can have 20 signal events, for 
values of $M_2$ upto about 200 GeV. Figure 3, on the higher hand,
shows the corresponding number of events to be around 50. 
In terms of neutralino chargino masses, this means that
we can probe upto a lighter chargino and second lightest neutralino
mass upto about 200 GeV. The lightest chargino mass reach is nearly
half of the above value. This is indicative of how much
improvement over the LEP results is possible through the VBF technique 
in probing  the chargino-neutrino sector of an R-parity violating
scenario.

The signals suggested here  can also be mimicked by the production of
charged-higgs boson pairs via VBF channel \cite{moretti}. The production
cross-section for the latter can reach upto  10 $fb$. 
Further decay of $H^\pm$ to $\chi_1^\pm \chi_1^0$ or 
$W^\pm h^0$, $tb$ may lead to final states comprising unlike-sign
dileptons associated with jets. Following reference 
\cite{mona_moretti}, charged-higgs branching ratio to $\chi_1^\pm\chi^0$ 
can be as big as 0.7. Leptonic decay of the charginos 
would make the effective branching ratio of a charged-higgs pair to
unlike sign dileptons around $2\%$. Cuts on the leptons and jets in the
central region will further reduce the signal strength and we may be
left with a maximum of 10 such events coming from this
channel. Moreover, R-conserving decays of the charginos would produce a
large amount of missing $E_T$ caused by the lightest neutralino. So our
missing $E_T$ cut ($<$ 10 GeV) can control this `background' quite
efficiently. 

Also, $H^\pm \rightarrow W^\pm h^0$ decay can have a
maximum branching ratio of 0.6 for low $\tan \beta$ and for higgs mass
less than $tb$ threshold. The $tb$ channel, once allowed, becomes 
dominant. Effective branching ratios of these channels to
unlike sign di-leptons are around 0.025 and 0.04 respectively. Both
these channels will produce missing energy due to the leptonic decays
of the $W$ bosons. Again our missing $E_T$ cut is effective in reducing 
these backgrounds compared to our signals. Finally, the direct
$H^\pm$ pair-production via $q \bar q$ or $gg$ fusion has a large
cross-section. Gluon radiation from initial states in such cases
may produce two forward jets. But demanding a high invariant mass 
of the forward jets and $E_T^j >$ 15 GeV, we can successfully kill this
background also.

\section{Concluding remarks}

We have considered the production of charginos and neutralinos at the
LHC via VBF in an R-parity violating scenario. We have seen that over
a substantial part of the parameter space, a $\lambda'$-type
interaction can cause the $\chi_2^{0}$ or the $\chi_1^{\pm}$ to decay
predominantly into a lepton and two quarks, thereby suppressing the
hadronically quiet trilepton mode and giving instead signals which can
be faked by gluino and squark cascades.  In these cases, charginos and
neutralinos, produced in pairs through VBF, can give rise to large
event rates, after filtration through cuts that should remove standard
model backgrounds and residual effects of the strongly interacting
sector of the SUSY spectrum.  We have also noted that the signals
suggested here can be clearly detectable after all cuts even for very
small values of the R-parity violating couplings.

In the context of (R-parity conserving) MSSM, too, the VBF channel
could be interesting. For example, VBF provides the unique channel for
the production of a $\chi_i^{\pm} \chi_j^{\pm}$ pair.  This will lead
to dileptons plus missing $E_T$ in the central region.  However, the
like-sign chargino pair-production cross-section in this manner is
rather low ($\sim 5-7 ~fb$) and the corresponding signals are
considerably smaller than those coming from the VBF production of
like-sign W's. On the other hand, unlike-sign dilepton event rates
from $\chi_i^{+} \chi_j^{-}$ production in the same manner fare much
better compared to the corresponding rate from $W^{+}W^{-}$
production.  For example, with $\mu~=~400~GeV,\; M_2~=~150~GeV$ and
$\tan \beta ~=~ 5$, and with the same cuts on the forward jets and
central leptons as those described in the previous section, $191$
{\em unlike-sign dileptons} $+ \; \miset$ events are predicted from a
$\chi_i^{+} \chi_j^{-}$ pair produced via VBF for an integrated
luminosity of $100~fb^{-1}$, as against $86$ from a $W$-pair produced
in the same fashion.  Since forward jet tagging is going to be a part
of LHC experiments anyway, this might be looked upon as an additional
process that can be exploited to uncover hadronically quiet signals of
the chargino-neutralino sector in MSSM.

Also, in cases with $\lambda$-type
interactions, although the R-violating decays of
$\chi_2^{0}$ or  $\chi_1^{\pm}$ will still give
hadronically quiet events, the VBF channels will
lead to spectacular multilepton
signals in the central region. Such central multilepton
signals can also
arise from the cascade process mentioned in the previous section when the
second lightest neutralino or the lighter chargino decays leptonically
into the lightest neutralino.
Similarly, decays of $\chi_2^{0}$ or  $\chi_1^{\pm}$ into real
gauge bosons and leptons might be the sources of interesting signals
of the bilinear R-parity violating term shown in equation (1). A detailed
study of such signals (including those within the MSSM) will be presented
in a subsequent paper \cite{vbf_susy}  .

Finally we want to point out that signals similar to those described 
here can also arise from the production of the lightest neutral Higgs 
boson and its subsequent decay into a pair of neutralinos which
might decay via the $\lambda'$-type couplings \cite{earlier_work}.
In such a case, however, the total invariant mass of the products
lying in the region in between the two forward jets is equal to the
Higgs mass, and should thus lie within about $130~GeV$. On the other 
hand, the signals from the chargino-neutralino pair have considerably
higher invariant mass so long as the chargino and neutralino masses 
satisfy the LEP constraints. Thus the signals discussed here are
in a way complementary to the ones that can be observed if the lightest
neutral scalar has a substantial branching ratio for the
$\chi^0_1 \chi^0_1$ channel.

{\bf Acknowledgement:} We thank D. Choudhury for helpful discussions.
The work of B.M. has been partialy supported by the Board of
Research in Nuclear Sciences, Department
of Atomic Energy, Government of India.
\newcommand{\plb}[3]{{Phys. Lett.} {\bf B#1} #2 (#3)}                  %
\newcommand{\prl}[3]{Phys. Rev. Lett. {\bf #1} #2 (#3) }        %
\newcommand{\rmp}[3]{Rev. Mod.  Phys. {\bf #1} #2 (#3)}             %
\newcommand{\prep}[3]{Phys. Rep. {\bf #1} #2 (#3)}                   %
\newcommand{\rpp}[3]{Rep. Prog. Phys. {\bf #1} #2 (#3)}             %
\newcommand{\prd}[3]{Phys. Rev. {\bf D#1} #2 (#3)}                    %
\newcommand{\np}[3]{Nucl. Phys. {\bf B#1} #2 (#3)}                     %
\newcommand{\npbps}[3]{Nucl. Phys. B (Proc. Suppl.)
           {\bf #1} #2 (#3)}                                           %
\newcommand{\sci}[3]{Science {\bf #1} #2 (#3)}                 %
\newcommand{\zp}[3]{Z.~Phys. C{\bf#1} #2 (#3)}                 %
\newcommand{\mpla}[3]{Mod. Phys. Lett. {\bf A#1} #2 (#3)}             %
 \newcommand{\apj}[3]{ Astrophys. J.\/ {\bf #1} #2 (#3)}       %
\newcommand{\jhep}[2]{{Jour. High Energy Phys.\/} {\bf #1} (#2) }%
\newcommand{\astropp}[3]{Astropart. Phys. {\bf #1} #2 (#3)}            %
\newcommand{\ib}[3]{{ ibid.\/} {\bf #1} #2 (#3)}                    %
\newcommand{\nat}[3]{Nature (London) {\bf #1} #2 (#3)}         %
 \newcommand{\app}[3]{{ Acta Phys. Polon.   B\/}{\bf #1} #2 (#3)}%
\newcommand{\nuovocim}[3]{Nuovo Cim. {\bf C#1} #2 (#3)}         %
\newcommand{\yadfiz}[4]{Yad. Fiz. {\bf #1} #2 (#3);             %
Sov. J. Nucl.  Phys. {\bf #1} #3 (#4)]}               %
\newcommand{\jetp}[6]{{Zh. Eksp. Teor. Fiz.\/} {\bf #1} (#3) #2;
           {JETP } {\bf #4} (#6) #5}%
\newcommand{\philt}[3]{Phil. Trans. Roy. Soc. London A {\bf #1} #2
        (#3)}                                                          %
\newcommand{\hepph}[1]{(hep--ph/#1)}           %
\newcommand{\hepex}[1]{(hep--ex/#1)}           %
\newcommand{\astro}[1]{(astro--ph/#1)}         %

\end{document}